\documentstyle[prb,twocolumn,aps]{revtex}

\input{epsf}

\begin{document}
\tolerance 10000

\draft

\title{Magnetic Induction of $d_{x^2 - y^2} + i \; d_{xy}$ Order
       in High-T$_c$ Superconductors}

\author{R. B. Laughlin}

\address{
Department of Physics\\
Stanford University\\
Stanford, California  94305 }

\twocolumn[
\date{ \today }
\maketitle
\widetext

\vspace*{-1.0truecm}

\begin{abstract}
\begin{center}
\parbox{14cm}{I propose that the phase transition in
Bi$_2$Sr$_2$CaCu$_2$O$_8$ recently observed by by Krishana et al
[Science {\bf 277}, 83 (1997)] is the development of a small $d_{xy}$
superconducting order parameter phased by $\pi/2$ with respect to the
principal $d_{x^2 - y^2}$ one to produce a minimum energy gap $\Delta$.
The violation of both parity and time-reversal symmetry allows the
development of a magnetic moment, the key to explaining the experiment.
The origin of this moment is a quantized boundary current of $I_B = 2
e \Delta / h$ at zero temperature.}
\end{center}
\end{abstract}

\pacs{
\hspace{1.9cm}
}
]

\narrowtext

In a recent paper Krishana et al \cite{ong} have reported a phase
transition in Bi$_2$Sr$_2$CaCu$_2$O$_8$ induced by a magnetic
field and characterized by a kink in the thermal conductivity
as a function of field strength, followed by a flat plateau.
The high-field state is also superconducting.  They argued from the
existence of this plateau that heat transport by quasiparticles was
zero in the new state and that this probably indicated the development
of an energy gap. The transition has the peculiarity
of being easily induced by small fields.  Krishana et al report the
empirical relation $T_c \propto \sqrt{B}$, although over the
limit field range of $0.6 T < B < 5 T$, and also that the transition
sharpens as $T_c$ is reduced.

I propose that the new high-field state is the parity and
time-reversal symmetry violating $d_{x^2 - y^2} + i d_{xy}$
superconducting state proposed long ago by me \cite{rbl}, which has
many properties in common with quantum hall states, including
particularly chiral edge modes and exactly quantized boundary currents.
The essential point of my argument is that the state must have a
magnetic moment in order to account for the experiment, and this is
possible only if it violates both parity and time-reversal symmetry.
The development of $s + id$ order, for example, or high-momentum Cooper
pairing \cite{ogata} are both ruled out for this reason, as is a
restructuring of the vortex lattice.  My hypothesis leads, through
reasoning described below, to the model free-energy functional

\begin{displaymath}
\frac{F}{L^2} = \frac{1}{6\pi} \frac{\Delta^3}{(\hbar v)^2}
- \frac{1}{\pi} \frac{eB}{\hbar c} \Delta \tanh^2(\frac{\beta \Delta}{2})
- \frac{4}{\pi} \frac{(k_B T)^3}{(\hbar v)^2}
\end{displaymath}

\begin{equation}
\times \biggl\{\frac{(\beta \Delta)^2}{2}
\ln [ 1 +  e^{-\beta \Delta} ] + \int_{\beta \Delta}^\infty
\ln [ 1 + e^{-x} ] x dx \biggr\} \; \; \; ,
\end{equation}

\noindent
where $\Delta$ is the induced energy gap and $v = \sqrt{v_2 v_2}$ is
the root-mean-square velocity of the d-wave node.  The value of the
latter is fixed by experiment, in particular photoemission bandwidth
\cite{photo} and the temperature dependence of the penetration depth
in YBCO \cite{hardy,bonn}.  Following Lee and Wen \cite{lee} I shall
use the values $v_1 = 1.18 \times 10^7$ cm/sec and $v_1/v_2 = 6.8$,
or $\hbar v = 0.30$ eV \AA.  The uncertainty in this number
is about 10\%.  At zero temperature the free energy is minimized by

\begin{equation}
\Delta_0 = \hbar v \sqrt{2 \frac{eB}{\hbar c}} \; \; \; ,
\end{equation}

\noindent
and has the value

\begin{equation}
\frac{F_0}{L^2} = - \frac{1}{3\pi} \frac{\Delta_0^3}{(\hbar v)^2}
\; \; \; .
\end{equation}

\begin{figure}
\epsfbox{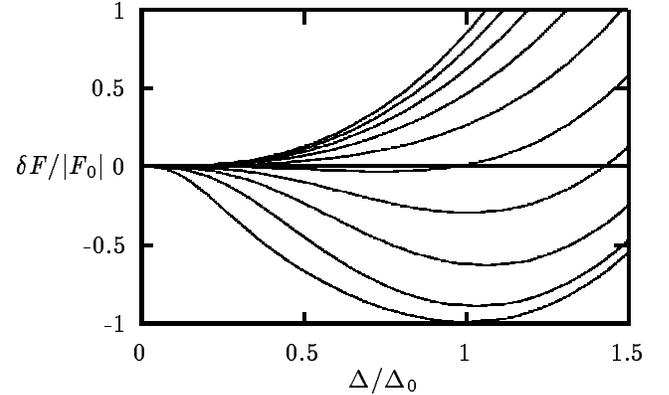}
\caption{Free energy versus $\Delta$ as described by Eq. (4) for
         temperatures $k_B T / \Delta_0 = 0.1 , 0.2 , ... , 1$.}
\end{figure}

\noindent
From the dimensionless version of this free energy

\begin{displaymath}
\frac{\delta F}{|F_0|} = \frac{1}{2} (\frac{\Delta}{\Delta_0})^3
- \frac{3}{2} (\frac{\Delta}{\Delta_0})
\tanh^2 (\frac{\beta \Delta}{2})+12 (\frac{k_B T}{\Delta_0})^3
\end{displaymath}

\begin{equation}
\times \biggl\{\int_0^{\beta \Delta} \ln [ 1 + e^{-x} ] x dx
- \frac{(\beta \Delta)^2}{2} \ln [ 1 +  e^{-\beta \Delta} ] \biggr\}
\end{equation}

\noindent
plotted in Fig. 1 I find a weakly first-order transition with

\begin{equation}
k_B T_c = 0.52 \; \Delta_0 \; \; \; .
\end{equation}

\noindent
This is plotted against the experiment in Fig. 2.  It will be seen
to account for both the functional form of the transition temperature
and its absolute magnitude with no adjustable parameters.

\begin{figure}
\epsfbox{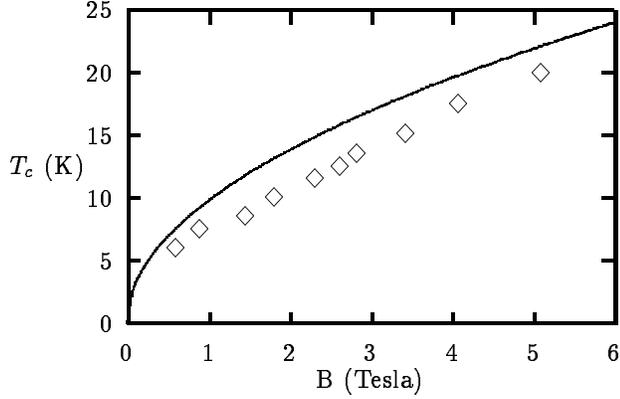}
\caption{Comparison of measured transition temperature versus magnetic
         field (diamonds) with Eq. (5).}
\end{figure}

There are three key steps leading Eq. (1):

\begin{enumerate}

\item The adoption of conventional quasiparticles at four nodes as the
      low-energy exitation spectrum of the parent $d_{x^2 - y^2}$
      state.

\item The derivation of a relation between the minimum energy to inject
      a quasiparticle in the bulk interior and a quantum-mechanical
      boundary current.

\item A guess as to the temperature dependence of this boundary
      current based on legitimate but model-dependent calculations.

\end{enumerate}

\noindent
The last of these, which I shall defend below, is pure phenomenology,
and I therefore consider the identification of the energy scale
$\Delta_0$ to be more significant than the first-orderness of the
transition or the specific factor 0.52 in Eq. (5).  The most important
matter is the quantization of the moment, which involves a conceptual
link between the $d_{x^2 - y^2} + i d_{xy}$ superconductivity and the
quantum hall effect, and which raises the possibility of new phenomena
at cuprate superconductor edges.

The assumption of conventional quasiparticles at d-wave nodes leads
to the repulsive $\Delta^3$ and free-quasiparticle entropy terms in
Eq. (1).  The model here is not critical, since only the node matters,
so let us use the BCS Hamiltonian

\begin{equation}
{\cal H} = \sum_{k s} \varepsilon_k \; c_{k s}^\dagger c_{ks}
+ \sum_{k k'} V_{k k'}
\; c_{k \uparrow}^\dagger c_{-k \downarrow}^\dagger
c_{-k' \downarrow} c_{k' \uparrow} \; \; \; .
\end{equation}

\noindent
As usual we consider variational ground states of the form

\begin{displaymath}
| \Psi \! > = \prod_k \biggl\{ u_k + v_k c_{k \uparrow}^\dagger
c_{-k \downarrow}^\dagger \biggr\} | 0 \! >
\end{displaymath}

\begin{equation}
|u_k |^2 + |v_k |^2 = 1 \; \; \; ,
\end{equation}

\noindent
and minimize the expected energy

\begin{displaymath}
< \! \Psi | {\cal H} | \Psi \! >
\end{displaymath}

\begin{equation}
= 2 \sum_k \varepsilon_k |v_k |^2
+ \sum_{k k'} V_{k k'} (u_k v_k^* ) (u_{k'}^* v_{k'})
\end{equation}

\noindent
to obtain

\begin{displaymath}
u_k = \sqrt{\frac{1}{2} \biggl[ 1 + \frac{\varepsilon_k}
{\sqrt{\varepsilon_k^2 + |\Delta_k^2|}} \biggr]}
\end{displaymath}

\begin{equation}
v_k = \sqrt{\frac{1}{2} \biggl[ 1 - \frac{\varepsilon_k}
{\sqrt{\varepsilon_k^2 + |\Delta_k^2|}} \biggr]} \frac{\Delta_k}
{| \Delta_k | }
\; \; \; \; ,
\label{d2}
\end{equation}

\noindent
where

\begin{equation}
\Delta_k = \sum_{k'} V_{k k'} (u_{k'}^* v_{k'}) \; \; \; ,
\end{equation}

\noindent
or

\begin{equation}
\Delta_k = - \frac{1}{2} \sum_{k'} V_{k k'} \frac{\Delta_{k'}}
{\sqrt{\varepsilon_{k'}^2 + | \Delta_{k'} |^2}} \; \; \; .
\label{gap}
\end{equation}

\noindent
Equivalently we may take Eqs. (\ref{d2}) to define $| \Psi \! >$ in
terms of $\Delta_k$ and minimize the expected energy

\begin{displaymath}
< \! \Psi | {\cal H} | \Psi \! > =
\sum_k \varepsilon_k \biggl[ 1 -  \frac{\varepsilon_k}
{\sqrt{\varepsilon_k^2 + |\Delta_k |^2}} \biggr]
\end{displaymath}

\begin{equation}
+ \frac{1}{4} \sum_{k k'} V_{k k'} \biggl[ \frac{\Delta_k^*}
{\sqrt{\varepsilon_k^2 + |\Delta_k |^2}} \biggr] \biggl[
\frac{\Delta_{k'}} {\sqrt{\varepsilon_{k'}^2 + |\Delta_{k'}
|^2}} \biggr] \; \; \; ,
\end{equation}

\noindent
to obtain Eq. (\ref{gap}).  Regardless of whether the extremal
condition is met the expected energy of the quasiparticle

\begin{equation}
| \Psi_{k \uparrow} \! > =
(u_k^* c_{k\uparrow}^\dagger + v_k^* c_{-k \downarrow} ) | \Psi \! >
\end{equation}

\noindent
is

\begin{equation}
< \! \Psi_{k \uparrow} | {\cal H} | \Psi_{k \uparrow} \! >
= < \! \Psi | {\cal H} | \Psi \! > + \sqrt{\varepsilon_k^2 +
| \Delta_k |^2} \; \; \; .
\end{equation}

\noindent
The prototypical $d_{x^2 - y^2} + i d_{xy}$ state is

\begin{equation}
\varepsilon_k = -2t \biggl[ \cos(k_x b) + \cos(k_y b) \biggr]
\end{equation}

\begin{displaymath}
\Delta_k = \Delta_{x^2 - y^2} \biggl[ \cos(k_x b) - \cos(k_y b) \biggr]
\end{displaymath}

\begin{equation}
+ i \Delta_{xy} \sin(k_x b) \sin(k_y b) \; \; \; .
\end{equation}

\noindent
The velocity in Eq. (1) is related to the model parameters by

\begin{equation}
\hbar v_1 = \sqrt{8} tb
\; \; \; \; \; \;
\hbar v_2 = \sqrt{2} \Delta_{x^2 - y^2} b
\; \; \; \; \; \;
v = \sqrt{v_1 v_2} \; \; \; .
\end{equation}

\noindent
Assuming now that the extremal condition requires $\Delta_{xy}$ to be
zero, so that the native ground state has only $d_{x^2 - y^2}$ order,
and then {\it forcing} the minimum quasiparticle energy to be
$\Delta$, one finds that the energy is minimized when

\begin{equation}
\Delta_{xy}^2
= \left[ \begin{array}{cl}
   \Delta^2 - (q/\hbar v)^2    & ; \; q \leq \Delta/\hbar v \\
   0                           & ; \; q > \Delta/\hbar v
\end{array} \right] \; \; \; ,
\label{dxy}
\end{equation}

\noindent
where q is the distance to the node in symmetrized units, and equals

\begin{displaymath}
\delta < \! \Psi | {\cal H} | \Psi \! > =
- \sum_k ( \varepsilon_k^2 + | \Delta_k |^2 ) \delta \biggl[
\frac{1}{\sqrt{\varepsilon_k^2 + | \Delta_k |^2}} \biggr]
\end{displaymath}

\begin{equation}
= \frac{2}{\pi} \hbar v L^2 \int_0^{\Delta/\hbar v} \biggl[
\frac{1}{q} - \frac{\hbar v}{\Delta} \biggr] q^3 dq
= \frac{L^2}{6\pi} \frac{\Delta^3}
{(\hbar v)^2} \; \; \; .
\end{equation}

\noindent
The quasiparticle contribution to the finite-temperature free energy
under these circumstances is

\begin{displaymath}
\frac{F_{{\rm quasi}}}{L^2} = - \frac{4}{\pi} \; k_B T
\end{displaymath}

\begin{equation}
\times \int_0^\infty \ln [ 1 +
\exp (- \beta \sqrt{ (\hbar q)^2 + \Delta_{xy}^2})]
q dq \; \; \; .
\end{equation}

Let us next consider the zero-temperature magnetic moment, which is
due to a circulating boundary current of

\begin{equation}
I_B = 2 \frac{e}{h} \Delta_0 \; \; \; .
\label{hall}
\end{equation}

\noindent
This works out to 0.13 $\mu$A for a gap of 1.64 meV induced by a
field 1 Tesla.  Boundary currents of this magnitude are known
to result from the development of a T-violating order parameter of
this size \cite{sauls}, so the issue is not the existence of these
currents or their disappearance when the second order parameter
vanishes but rather their specific functional dependence on $\Delta$
and sense of circulation.  T {\it and} P must both be violated for
the boundary currents to generate a moment. The $s + id$ state, for
example, will not work because its reflection symmetry about the x-axis
forces the currents at the +y and -y edges to flow in the same
direction, whereas flow in opposite directions is required to generate
a moment.  The functional form comes from the ability to continuously
deform a $d_{x^2 - y^2} + i d_{xy}$ state into a quantum hall state
without closing the gap.  This is demonstrated with the following
simple lattice example.  Let

\begin{displaymath}
{\cal H}_{HF} = 2t \sum_k \biggl\{ \biggl[\cos (k_x b) + \cos (k_y b)
\biggr] \Psi_k^\dagger \tau_3 \Psi_k
\end{displaymath}

\begin{displaymath}
+ \biggl[ \cos (k_x b) - \cos (k_y b) \biggr] \Psi_k^\dagger\tau_1
\Psi_k
\end{displaymath}

\begin{equation}
+ 2 m \sin(k_x ) \sin (k_y ) \Psi_k^\dagger \tau_2 \Psi_k \biggr\}
\label{flux}
\end{equation}

\noindent
be the Hartree-Fock Hamiltonian for a $d_{x^2 - y^2} + i d_{xy}$
superconductor on a square lattice, where $m$ is a constant
characterizing the size of the energy gap and

\begin{displaymath}
\Psi_k = \left[ \begin{array}{c}
c_{k \uparrow} \\ c_{-k \downarrow}^\dagger
\end{array} \right]
\end{displaymath}

\begin{equation}
\tau_1 =
\left[ \begin{array}{lr}
0 & \; 1 \\  1 & 0
\end{array} \right]
\; \; \; \;
\tau_2 =
\left[ \begin{array}{lr}
0 & \! -i \\  i & 0
\end{array} \right]
\; \; \; \;
\tau_3 =
\left[ \begin{array}{lr}
 1 & 0 \\  0 & \! -1
\end{array} \right]
\; \; \; ,
\end{equation}

\noindent
as usual. Let the j$^{th}$ lattice site be denoted

\begin{equation}
\vec{r}_j = \left[ \begin{array}{c}
\ell_j \\ m_j \end{array} \right] b \; \; \; .
\end{equation}

\noindent
Then the Hamiltonian $U(\theta) {\cal H}_{HF} U^\dagger(\theta)$, where

\begin{displaymath}
U(\theta) = \exp \biggl\{\frac{\theta}{2} \sum_j
\end{displaymath}

\begin{equation}
\biggl[ 1 - 2 (-1)^{m_j} + (-1)^{\ell_j + m_j} \biggr]
( c_{j \uparrow}^\dagger c_{j \downarrow}^\dagger -
c_{j \downarrow} c_{j \uparrow} ) \biggr\} \; \; \; ,
\end{equation}

\noindent
interpolates between ${\cal H}_{HF}$ at $\theta = 0$ and a lattice
Landau level Hamiltonian at $\theta = \pi/4$, all the while having the
same eigenvalue spectrum due to the unitarity of $U(\theta)$
\cite{kotliar,zou}.  Because of the existence of this mapping it is
possible to perform thought experiments, such as that illustrated in
Fig. 3, in which flux is wound adiabatically through a loop of
superconductor to which quantum hall states are attached as electrodes.
The notion of adiabatic in this case is somewhat more subtle than in
the traditional quantum hall argument because of the broken symmetry
in the sample interior, but the end result is the same \cite{rbl1}:
Slow insertion of a 1-particle flux quantum $hc/e$ through the loop
causes spectral flow of {\it two} quasiparticle states from bulk
interior to the edge where they are deposited at the chemical potential.
The energy increase associated with this transfer is $2 \Delta$, where
$\Delta$ is the energy required to inject a quasiparticle into its
lowest-energy state in the bulk interior. Only the lowest-energy state
matters because the anti-crossing rule prevents any higher bulk states
from flowing to the chemical potential. Flux addition also induces bulk
supercurrent, but this is easily removed by causing the ring
circumference to $L$ diverge, since the energy in question is

\begin{equation}
\delta E_{{\rm bulk}} = \frac{\hbar^2}{2m} (\frac{2\pi}{L})^2
\int n_s (\vec{r} ) d\vec{r} \; \; \; ,
\end{equation}

\noindent
where $n_s$ is the superfluid density, which falls off as 1/L.
Equivalently one may say that there is is a physical difference
between current already present and current induced by the injected
flux. In any case the boundary current is c times the energy
increase $2\Delta_0$ divided by the flux quantum $hc/e$, which gives
Eq. (\ref{hall}).  Thus at zero temperature the $d_{x^2 - y^2} + i
d_{xy}$ superconductor possesses a quantized hall conductance in which
the gradient of the electrostatic potential is replaced by the gradient
of the gap. This result is exact.

\begin{figure}
\epsfbox{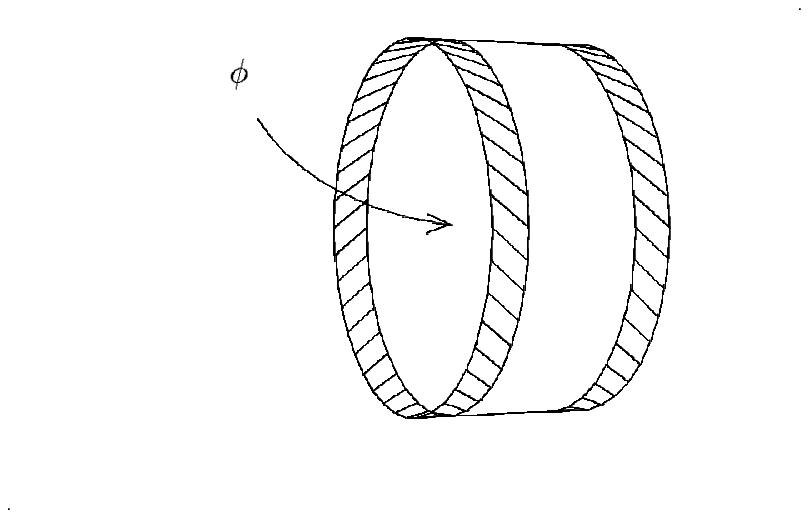}
\caption{Illustration of thought experiment in which quantum hall
states are used as left and right electrical contacts for a
$d_{x^2 - y^2} + i d_{xy}$ superconductor.}
\end{figure}

The final matter for consideration is the reduction of this moment
by thermal excitation of quasiparticles.  This is, unfortunately,
sensitive to details and thus difficult to calculate with
sufficient accuracy to describe the phase transition.  It can be
understood simply in terms of the flux Hamiltonian obtained by
rotating Eq. (\ref{flux}) by $\theta = \pi/4$.  This consists of
upper and lower quantum hall bands with opposite quantizations,
these being manifested primarily in the states of energy near
$\Delta = 4tm$. Evaluating the Hall conductance of this model by
the Kubo formula in the limit of small $m$ we find that \cite{zou}

\begin{equation}
\sigma_{xy} = \frac{e^2}{h} \int_0^\infty \tanh \biggl[
\frac{\beta \Delta}{2} \sqrt{1 + x} \biggr] \frac{dx}{(1 + x)^{3/2}}
\; \; \; .
\end{equation}

\noindent
The strong quenching effect at $k_B T \geq \Delta$ occurs because
free quasiparticles contribute a Hall conductance opposite to that of
the ground state. Assuming now that $\Delta$ varies slowly in space and
equals zero at the sample edge, we may integrate in from the edge to
obtain

\begin{equation}
I_B \simeq 2 \frac{e}{h} k_B T \int_0^\infty \ln \biggl[ \cosh (\frac{
\beta \Delta}{2}\sqrt{1 + x}) \biggr] \frac{dx}{(1 + x)^2}
\; \; \; .
\end{equation}

\noindent
The version of this appropriate to Eq. (\ref{dxy}) is

\begin{equation}
I_B \simeq 4 \frac{e}{h} k_B T \ln \biggl[ \cosh (\frac{
\beta \Delta}{2} ) \biggr] \; \; \; .
\end{equation}

\noindent
This is quite close to the functional form appearing in Eq. (1) in the
range of interest, saturates to linearity in $\Delta$ at zero
temperature, becomes exponentially quenched for temperatures
$k_B T >> \Delta$, but gives no phase transition. The
phenomenological function I chose is merely an approximation to this
one constrained to be analytic in $\Delta$ and odd.  The
proportionality of $T_c$ to $\Delta_0$, however, is expected on general
grounds because there is no other energy scale in the problem.

The complete absence of thermal transport above $T_c$ in the experiment
is not explained by thermal activation to a gap of order $\Delta_0$,
as this is simply too small to freeze out all the quasiparticles.
This criticism, however, applies equally well to any theory of the
effect one would care to consider, for it is physically unreasonable
for a gap much larger than $k_B T_c$ to develop spontaneously.  I
therefore believe that absence of transport is an effect of enhanced
scattering and trapping of quasiparticles in the new state and is a
detail to be worked out once the symmetry of the second order
parameter is established.  There is certainly the potential for
violent scattering in the $d_{x^2 - y^2} + i d_{xy}$ state given the
inhomogeneity of the magnetic field due to the vortex lattice and the
possibility that the transition is weakly first-order, but it is a
mistake to use this as a criterion for deciding whether the symmetry
I have identified is the right one.

I wish to express special thanks to C. M. Varma for alerting me to
the large moment carried by this class of superconductor, and to
N. P. Ong, F. D. M. Haldane, J. Berlinsky, C. Kallin, and A. Balatsky
for helpful discussion and criticism. This work was supported primarily
by the NSF under grant No. DMR-9421888.  Additional support was
provided by the Center for Materials Research at Stanford University
and by NASA Collaborative Agreement NCC 2-794.

\end{document}